\documentclass[twocolumn,american]{revtex4-2}
\usepackage[T1]{fontenc}
\usepackage[utf8]{luainputenc}
\usepackage{amsmath}
\usepackage{babel}
\begin{document}
\title{Classical action for the height within the Kardar-Parisi-Zhang (KPZ)
equation for surface growth}
\author{Garry Goldstein}
\address{garrygoldsteinwinnipeg@gmail.com}
\begin{abstract}
In this work we write down a classical (not quantum) action for the
surface height for the Kardar-Parisi-Zhang (KPZ) equation for surface
growth. We do so starting with the regular Martin-Siggia-Rose (MSR)
action (which is quantum - contains the constraint field) and integrate
out the quantum constraint field exactly. We analyze the classical
action, we thereby obtain, within the gaussian and one loop approximations
various instabilities to rough surfaces. The gaussian analysis predicts
instabilities to rough surfaces below two dimensions while one loop
analysis predicts stronger surface stability to roughness and instabilities
to roughness below one dimension rather then two dimensions. The one
loop analysis shows that the KPZ action is incomplete and that we
generate additional terms (not found in the initial classical action)
in the action perturbatively. In the supplement we also modify the
KPZ equation for growing surfaces to include the effect of surface
tilt on the noise (that is have two sources of noise one of which
is multiplicatively coupled noise).
\end{abstract}
\maketitle

\section{ Introduction}\label{sec:Introduction}

An important example of a non-equilibrium statistical mechanics system
is a surface growing under the external deposition of molecules. A
mathematical, stochastic differential, equation that describes a very
wide class of such growing surfaces is known as the Kardar-Parisi-Zhang
(KPZ) equation. It is given by \citep{Kardar_1986,Kamenev_2011,Forster_1977}:
\begin{align}
\frac{\partial h}{\partial t} & =\Gamma\left(\sqrt{1+\left(\nabla h\right)^{2}}\right)-D\nabla^{2}h+\xi\nonumber \\
 & \cong\Gamma\left(1+\frac{1}{2}\left(\nabla h\right)^{2}\right)-D\nabla^{2}h+\xi.\label{eq:KPZ_old}
\end{align}
Here $h$ is the surface height, $D$ is the diffusion coefficient
and $\Gamma$ is an overall rate of molecule deposition and we have
performed a small gradient expansion for simplicity. Furthermore $\xi\left(\mathbf{x},t\right)$
is a Gaussian noise function with action:
\begin{equation}
\int\mathcal{D}\xi\exp\left(-\int d^{d}\mathbf{x}dt\xi^{2}/2\Delta\right)\label{eq:Action}
\end{equation}
Also the second term on the RHS of Eq. (\ref{eq:KPZ_old}) describes
the diffusion of atoms on the surface which tends to smooth it out.
Furthermore for geometric reasons we see that the surface height growth
rate is not constant $\Gamma$ but is given by $\frac{\Gamma}{\cos\left(\theta\right)}$
where $\theta$ is the angle between the normal to the surface ($\hat{n}$)
and the direction the molecules are deposited - say z-axis.We now
compute the normal to a $2d$ surface in terms of the gradient $\frac{\partial h}{\partial x}$
and $\frac{\partial h}{\partial y}$ whereby we obtain: 
\begin{equation}
\hat{n}\sim\left(\begin{array}{ccc}
\hat{x} & \hat{y} & \hat{z}\\
1 & 0 & \frac{\partial h}{\partial x}\\
0 & 1 & \frac{\partial h}{\partial y}
\end{array}\right)=-\frac{\partial h}{\partial x}\hat{x}-\frac{\partial h}{\partial y}\hat{y}+\hat{z}\label{eq:Normal-1}
\end{equation}
Whereby: 
\begin{equation}
\Rightarrow\cos\left(\hat{n}\cdot\hat{z}\right)=\frac{1}{\sqrt{1+\left(\frac{\partial h}{\partial x}\right)^{2}+\left(\frac{\partial h}{\partial y}\right)^{2}}}\label{eq:Normal_angle-1}
\end{equation}
As such we obtain Eq. (\ref{eq:KPZ_old}). Originally the KPZ equation
was proposed by Kardar, Parisi and Zhang to study the growth of surfaces
\citep{Kardar_1986,Kamenev_2011} however it was later realized that
the same equations appear in a wide variety of contexts. For example
the KPZ equation arises in the weakly asymmetric simple exclusion
principle dynamics \citep{Beretini_1997}, in the partition functions
of directed polymers \citep{Brunet_2000,Wiese_2022}, Burgers' hydrodynamics
(which Navier-Stokes hydrodynamics without vorticity and with noise)
\citep{Kamenev_2011,Martin_1973,Wiese_2022}, polynuclear growth \citep{Prahofer_2000,Prahofer_2000-1,Prahofer_2002},
rigid solid on solid (RSOS) models \citep{Martini_2000,Martini_2002,Alves_2014,Gomes_2019}
and represents a whole universality class of processes with both space
and time dependence \citep{Corwin_2012,Wiese_2022}. 

The KPZ equation may be studied in a variety of field theoretical
ways. It can be studied using the quantum KPZ action (based on the
Martin-Siggia-Rose (MSR) formalism). The action for this approach
is given by \citep{Kamenev_2011}: 
\begin{equation}
S\left(h,p\right)=-\int d^{d}\mathbf{x}dt\left[p\frac{\partial h}{\partial t}-Dp\nabla^{2}h+\frac{1}{2}\Gamma p\left(\nabla h\right)^{2}-\Delta p^{2}\right]\label{eq:Quantum_action}
\end{equation}
Here $p$ is the quantum constraint field. Alternatively one can use
a Cole-Hopf transformation \citep{Wiese_2022,Kamenev_2011}, with
the transformation being given by:
\begin{equation}
\varphi=\exp\left(\frac{\Gamma}{2D}h-\frac{\Gamma^{2}\Delta}{4D^{2}}t\right)\label{eq:Cole_Hopf}
\end{equation}
Whereby using the Martin-Siggia-Rose (MSR) procedure one obtains the
action: 
\begin{equation}
S\left(\varphi,\pi\right)=-\int d^{d}\mathbf{x}dt\left[\pi\left(\frac{\partial\varphi}{\partial t}-D\nabla^{2}\varphi\right)+\frac{\Gamma^{2}\Delta}{4D^{2}}\pi^{2}\varphi^{2}\right]\label{eq:Cole_Hopf_action}
\end{equation}
Here $\pi$ is the quantum constraint field. For both approaches one
then obtains that for all $d\leq2$ (in two or less dimensions) the
action flows to a strong coupling fixed point (where $\frac{\Gamma^{2}\Delta}{4D^{2}}$
becomes large and perturbation theory fails) often associated with
rough surfaces.

In this work we study the KPZ equation using the MSR method, without
the Cole-Hopf transform, e.g. using the action in Eq. (\ref{eq:Quantum_action}).
We integrate out the quantum constraining field exactly (which can
be done at tree level as, after a co-ordinate change, there is only
one connected Feynman diagram relevant to the constraining field)
and transform the KPZ - MSR quantum - action into a classical action
just for the height field (see Eq. (\ref{eq:Action_height})). We
analyze the action in Eq. (\ref{eq:Action_height}) for instabilities
to rough surfaces using gaussian and one loop like analysis. We find
the upper critical dimension for rough surfaces is $d=2$ for the
gaussian analysis and $d=2$ is modified using the one loop analysis
with rough surfaces forming for $d<1$. We argue that the system is
partially amenable to Renormalization Group (RG) analysis and propose
new self energy terms and interaction terms that show up perturbatively
in the RG process (ones that do not show up in the original action
\citep{Continentino_2017,Sachdev_2011} but are more relevant in the
RG sense).

\section{ Main calculations}\label{sec:Main-manipulations}
\begin{widetext}
Consider the MSR action for the KPZ equation \citep{Kamenev_2011,Martin_1973}
given in Eq. (\ref{eq:Quantum_action}) (see also the supplementary
online information \citep{Supplement_2024}). Now we write:
\begin{equation}
-\Delta p^{2}+p\frac{\partial h}{\partial t}-Dp\nabla^{2}h=-\Delta\left[p-\frac{1}{2\Delta}\left(\frac{\partial h}{\partial t}-D\nabla^{2}h\right)\right]^{2}+\frac{1}{4\Delta}\left(\frac{\partial h}{\partial t}-D\nabla^{2}h\right)^{2}\label{eq:Manipulation_I}
\end{equation}
and
\begin{equation}
\frac{1}{2}\Gamma p\left(\nabla h\right)^{2}=\frac{1}{2}\Gamma\left[p-\frac{1}{2\Delta}\left(\frac{\partial h}{\partial t}-D\nabla^{2}h\right)\right]\left(\nabla h\right)^{2}+\frac{\Gamma}{4\Delta}\left(\frac{\partial h}{\partial t}-D\nabla^{2}h\right)\left(\nabla h\right)^{2}\label{eq:Manipulation_II}
\end{equation}
As such we have rewritten the action in Eq. (\ref{eq:Quantum_action})
in terms of two fields $h$ and $P=p-\frac{1}{2\Delta}\left(\frac{\partial h}{\partial t}-D\nabla^{2}h\right)$.
Now we write: 
\begin{align}
\mathcal{Z} & =\int\mathcal{D}h\mathcal{D}p\exp\left(-\int d^{d}\mathbf{x}dt\left[p\frac{\partial h}{\partial t}-Dp\nabla^{2}h-\Delta p^{2}+\frac{1}{2}\Gamma p\left(\nabla h\right)^{2}\right]\right)\nonumber \\
 & =\int\mathcal{D}h\mathcal{D}P\exp\left(-\int d^{d}\mathbf{x}dt\left[-\Delta P^{2}-\frac{1}{4\Delta}\left(\frac{\partial h}{\partial t}-D\nabla^{2}h\right)^{2}+\frac{1}{2}\Gamma P\left(\nabla h\right)^{2}+\frac{\Gamma}{4\Delta}\left(\frac{\partial h}{\partial t}-D\nabla^{2}h\right)\left(\nabla h\right)^{2}\right]\right)\nonumber \\
 & =\int\mathcal{D}h\exp\left(-\int d^{d}\mathbf{x}dt\left[\frac{1}{4\Delta}\left(\frac{\partial h}{\partial t}-D\nabla^{2}h\right)^{2}+\frac{\Gamma^{2}}{4\Delta}\left(\nabla h\right)^{4}+\frac{\Gamma}{4\Delta}\left(\frac{\partial h}{\partial t}-D\nabla^{2}h\right)\left(\nabla h\right)^{2}\right]\right)\label{eq:Path_integral_calculation}
\end{align}
where we have performed a change of variables $\mathcal{D}h\mathcal{D}p\rightarrow\mathcal{D}h\mathcal{D}P$.
We note that the determinant of the Jacobian of the transformation
is unity as the Jacobian of the transformation is upper triangular.
We have integrated out $P$ to tree level (which is exact as there
is only one connected Feynman diagram for $P$ and we use exponentiation
of disconnected diagrams \citep{Peskin_1995}) and we have introduced
the integral: 
\begin{equation}
\frac{\int\exp\left(\Delta P^{2}\right)P^{2}dP}{\int\exp\left(\Delta P^{2}\right)dP}=\frac{-\int\exp\left(-\Delta P^{2}\right)P^{2}d\left[iP\right]}{\int\exp\left(-\Delta P^{2}\right)d\left[iP\right]}=-\frac{1}{\Delta}\label{eq:Regularization}
\end{equation}
where the variable $P$ was Wick rotated as a regularization. As such
we have obtained a new KPZ action:

\begin{equation}
S\left(h\right)=-\int d^{d}\mathbf{x}dt\left[\frac{1}{4\Delta}\left(\frac{\partial h}{\partial t}-D\nabla^{2}h\right)^{2}+\frac{\Gamma}{4\Delta}\left(\frac{\partial h}{\partial t}-D\nabla^{2}h\right)\left(\nabla h\right)^{2}+\frac{\Gamma^{2}}{4\Delta}\left(\nabla h\right)^{4}\right]\label{eq:Action_height}
\end{equation}
This is the main result of this work as it integrates out the quantum
action for KPZ into a classical action just for the height field.
\end{widetext}

\section{Analysis of the action in Eq. (\ref{eq:Action_height})}\label{sec:Analysis-of-the}

We will study the properties of the action in Eq. (\ref{eq:Action_height}).
We will do gaussian and one loop analysis of the action. We will not
do Renormalization Group (RG) analysis due to current questions about
its accuracy \citep{Goldstein_2024} for obtaining fixed point positions
and coefficients and because preliminary one loop analysis shows that
RG will generate new RG relevant couplings not found in Eq. (\ref{eq:Action_height}).

\subsection{Gaussian Analysis}\label{subsec:Gaussian-Analysis}

We now Fourier transform to obtain: 
\begin{equation}
h\left(\mathbf{x},t\right)=\frac{1}{\left(2\pi\right)^{d+1}}\int d^{d}\mathbf{k}d\omega h\left(\mathbf{k},\omega\right)\exp\left(i\left[\mathbf{k}\cdot\mathbf{x}-\omega t\right]\right)\label{eq:Fourier_transform}
\end{equation}
Now lets first focus on the quadratic action from Eq. (\ref{eq:Action_height}):
\begin{equation}
S_{Quad}\left(h\right)=-\frac{1}{\left(2\pi\right)^{d+1}}\int d^{d}\mathbf{k}d\omega\left|h\left(\mathbf{k}\right)\right|^{2}\frac{1}{2\Delta}\left[\omega^{2}+D^{2}\mathbf{k}^{4}\right]\label{eq:Quadratic_action}
\end{equation}
This means that \citep{Auerbach_1994}:
\begin{align}
 & \left\langle \left[h\left(\mathbf{x},t\right)-h\left(\mathbf{0},0\right)\right]^{2}\right\rangle \nonumber \\
 & =\frac{2\Delta}{\left(2\pi\right)^{d+1}}\int d^{d}\mathbf{k}d\omega\frac{\left[1-\exp\left(i\left[\mathbf{k}\cdot\mathbf{x}-\omega t\right]\right)\right]}{\left(\omega^{2}+D^{2}\mathbf{k}^{4}\right)}\label{eq:Correlation}
\end{align}
We now perform this integral, we do the $\omega$ integration first,
close the contour in the lower half plane and pick up the pole at
$\omega=-iD\mathbf{k}^{2}$ to obtain: 
\begin{align}
 & \frac{1}{\left(2\pi\right)^{d+1}}\int d^{d}\mathbf{k}d\omega\frac{\left[1-\exp\left(i\left[\mathbf{k}\cdot\mathbf{x}-\omega t\right]\right)\right]}{\left(\omega^{2}+D^{2}\mathbf{k}^{4}\right)}\nonumber \\
 & =\frac{1}{\left(2\pi\right)^{d}}\int d^{d}\mathbf{k}\frac{1}{2D\mathbf{k}^{2}}\left[1-\exp\left(i\mathbf{k}\cdot\mathbf{x}-D\mathbf{k}^{2}t\right)\right]\nonumber \\
 & \cong\frac{V\left(S^{d-1}\right)}{2D\left(2\pi\right)^{d}}\left[\frac{1}{2-d}\left[a^{2-d}-L^{2-d}\right]-\int_{1/L}^{1/\sqrt{Dt+\left|\mathbf{x}\right|^{2}}}k^{d-3}dk\right]\nonumber \\
 & =\frac{V\left(S^{d-1}\right)}{\left(2-d\right)2D\left(2\pi\right)^{d}}\left[a^{2-d}-\sqrt{Dt+\left|\mathbf{x}\right|^{2}}^{2-d}\right]\label{eq:Integrals-1}
\end{align}
Here $V\left(S^{d-1}\right)$ is the volume of a unit radius $d-1$
dimensional sphere.We now see that for: 
\begin{equation}
\left\{ \begin{array}{cc}
d>2 & smooth\,surface\\
d=2 & marginal\,surface\\
d<2 & rough\,surface
\end{array}\right.\label{eq:conclusions}
\end{equation}
Indeed the term $a^{2-d}-\sqrt{Dt+\left|\mathbf{x}\right|^{2}}^{2-d}$
becomes large for $d<2$ for large distances and times showing non-smooth
surfaces. Here $a$is the typical size of a molecule which serves
as a cutoff and is considered finite, that is it is not a divergent
term in the calculations.

\subsection{ One loop analysis for self energy}\label{subsec:Meanfield-analysis}

We notice that there are only two one loop diagrams for self energy
$\Sigma\left(\mathbf{k},\omega\right)$. Let us only pick out terms
which are proportional to $\mathbf{k}^{2}$ and $\omega^{2}$ that
modify the dispersion (these are the most divergent term). We will
see that the terms $\sim\omega^{2}$ will be non-analytic but can
be regularized to be analytic in an RG way \citet{Sachdev_2011}.
At one loop there are two Feynman diagrams contributing diagrams with
their values given by: 
\begin{equation}
\Sigma\left(\mathbf{k},\omega\right)=\Sigma_{1}\left(\mathbf{k},\omega\right)+\Sigma_{2}\left(\mathbf{k},\omega\right)+\Sigma_{3}\left(\mathbf{k},\omega\right)+\Sigma_{4}\left(\mathbf{k},\omega\right)\label{eq:Sum}
\end{equation}
Where:

\begin{align}
\Sigma_{1}\left(\mathbf{k},\omega\right) & =2\Gamma^{2}\mathbf{k}^{2}\int\frac{d\Omega d^{d}\mathbf{q}}{\left(2\pi\right)^{d+1}}\frac{\mathbf{q}^{2}}{\Omega^{2}+D^{2}\mathbf{q}^{4}}\nonumber \\
 & =2\Gamma^{2}\mathbf{k}^{2}\int\frac{1}{\left(2\pi\right)^{d}}\frac{d^{d}\mathbf{q}}{2D\mathbf{q}^{2}}\mathbf{q^{2}}\nonumber \\
 & =\frac{\Gamma^{2}\mathbf{k}^{2}V\left(S^{d-1}\right)}{\left(2\pi\right)^{d}D\cdot d\cdot a^{d}}\label{eq:Self_energy}
\end{align}

\begin{align}
\Sigma_{2}\left(\mathbf{k},\omega\right) & =\frac{2}{d}\Gamma^{2}\mathbf{k}^{2}\int\frac{d\Omega d^{d}\mathbf{q}}{\left(2\pi\right)^{d+1}}\frac{\mathbf{q}^{2}}{\Omega^{2}+D^{2}\mathbf{q}^{4}}\nonumber \\
 & =\frac{2}{d}\Gamma^{2}\mathbf{k}^{2}\int\frac{1}{\left(2\pi\right)^{d}}\frac{d^{d}\mathbf{q}}{2D\mathbf{q}^{2}}\mathbf{q^{2}}\nonumber \\
 & =\frac{\Gamma^{2}\mathbf{k}^{2}V\left(S^{d-1}\right)}{\left(2\pi\right)^{d}D\cdot d^{2}\cdot a^{d}}\label{eq:Sigma_2}
\end{align}

\begin{align}
\Sigma_{3}\left(\mathbf{k},\omega\right) & =-\frac{2}{d}\Gamma^{2}\mathbf{k}^{2}\int\frac{d\Omega d^{d}\mathbf{q}}{\left(2\pi\right)^{d+1}}\frac{\mathbf{q}^{2}}{\Omega^{2}+D^{2}\mathbf{q}^{4}}\nonumber \\
 & =-\frac{2}{d}\Gamma^{2}\mathbf{k}^{2}\int\frac{1}{\left(2\pi\right)^{d}}\frac{d^{d}\mathbf{q}}{2D\mathbf{q}^{2}}\mathbf{q^{2}}\nonumber \\
 & =-\frac{\Gamma^{2}\mathbf{k}^{2}V\left(S^{d-1}\right)}{\left(2\pi\right)^{d}D\cdot d^{2}\cdot a^{d}}\label{eq:Sigma_3}
\end{align}
And: 
\begin{align}
\Sigma_{4}\left(\mathbf{k},\omega\right) & \cong-2\Gamma^{2}\omega^{2}\int\frac{d\Omega d^{d}\mathbf{q}}{\left(2\pi\right)^{d+1}}\frac{\mathbf{q}^{4}}{\left[\Omega^{2}+D^{2}\mathbf{q}^{4}\right]^{2}}\nonumber \\
 & =-2\Gamma^{2}\omega^{2}\int\frac{d\Omega d^{d}\mathbf{q}}{\left(2\pi\right)^{d+1}}\frac{\mathbf{q}^{4}}{\left[\Omega+iD\mathbf{q}^{2}\right]^{2}\left[\Omega-iD\mathbf{q}^{2}\right]^{2}}\nonumber \\
 & =-2\Gamma^{2}\omega^{2}\int\frac{d^{d}\mathbf{q}d\Omega}{\left(2\pi\right)^{d+1}}\frac{\mathbf{q}^{4}}{\Omega^{2}\left[\Omega+i2D\mathbf{q}^{2}\right]^{2}}\nonumber \\
 & =-2\Gamma^{2}\omega^{2}\int\frac{d\Omega d^{d}\mathbf{q}}{\left(2\pi\right)^{d+1}}\frac{\mathbf{q}^{4}}{-4\Omega^{2}D\mathbf{q}^{4}}\left[1-\frac{\Omega}{iD\mathbf{q^{2}}}+...\right]\nonumber \\
 & =-\frac{1}{2}\Gamma^{2}\omega^{2}\int\frac{d^{d}\mathbf{q}}{\left(2\pi\right)^{d}}\frac{1}{D^{2}\mathbf{q}^{2}}\label{eq:Unregulated}
\end{align}
Now we regularize the action by inserting the incoming momenta 
\begin{align}
\Rightarrow\Sigma_{4}\left(\mathbf{k},\omega\right)\cong & -\frac{1}{2}\Gamma^{2}\omega^{2}\int\frac{d^{d}\mathbf{q}}{\left(2\pi\right)^{d}}\frac{1}{D^{2}\left[\mathbf{q}^{2}+\left(\mathbf{k}^{2}+\omega^{2}\right)\right]}\nonumber \\
 & =-\frac{1}{2}\Gamma^{2}\omega^{2}\frac{V\left(S^{d-1}\right)}{\left(2\pi\right)^{d}D^{2}\cdot\left(d-2\right)\cdot a^{d-2}}\nonumber \\
 & \;+\frac{1}{2}\Gamma^{2}\omega^{2}\frac{V\left(S^{d-1}\right)\left(\mathbf{k}^{2}+\omega^{2}\right)^{d-2}}{\left(2\pi\right)^{d}D^{2}\cdot\left(d-2\right)}\label{eq:Regularized}
\end{align}
We now do RG regularization of the non-analytical terms (see \citep{Sachdev_2011}
section 3.3.2):
\begin{equation}
\Rightarrow\Sigma_{4}\left(\mathbf{k},\omega\right)\cong-\frac{1}{2}\Gamma^{2}\omega^{2}\frac{V\left(S^{d-1}\right)}{\left(2\pi\right)^{d}D^{2}\cdot\left(d-2\right)\cdot a^{d-2}}\label{eq:Slef_energy_final-1}
\end{equation}
Indeed RG would in introduce momentum shell integrals whereby the
lower cutoff in Eq. (\ref{eq:Regularized}) would no longer be introduced
and we have dropped a constant of order unity in the final answer
(this constant cannot be reliably computed see Ref. \citep{Goldstein_2024}).
We see that $d=2$ is the upper critical dimension for the self energy.
As such we may write
\begin{align}
 & \left\langle \left[h\left(\mathbf{x},t\right)-h\left(\mathbf{0},0\right)\right]^{2}\right\rangle \nonumber \\
 & =\frac{1}{\left(2\pi\right)^{d+1}}\int d^{d}\mathbf{k}d\omega\frac{\left[1-\exp\left(i\left[\mathbf{k}\cdot\mathbf{x}-\omega t\right]\right)\right]}{E\left(\mathbf{k},\omega\right)}\label{eq:Correlation-1}
\end{align}
With:
\begin{align}
E\left(\mathbf{k},\omega\right) & =\left[\frac{1}{2\Delta}-\frac{\Gamma^{2}V\left(S^{d-1}\right)}{2\left(2\pi\right)^{d}D^{2}\cdot\left(d-2\right)\cdot a^{d-2}}\right]\omega^{2}\nonumber \\
 & +\frac{\Gamma^{2}V\left(S^{d-1}\right)}{\left(2\pi\right)^{d}D\cdot d\cdot a^{d}}\mathbf{k}^{2}\label{eq:Slef_energy_final}
\end{align}
We now see that for: 
\begin{equation}
\left\{ \begin{array}{cc}
d>1 & smooth\,surface\\
d=1 & marginal\,surface\\
d<1 & rough\,surface
\end{array}\right.\label{eq:conclusions-1}
\end{equation}
This is similar to Eq. (\ref{eq:conclusions}). We note that without
RG regularization we would have obtained non-derivative couplings
similar to Hertz-Millis theory \citep{Hertz_1976,Millis_1993,Continentino_2017}.
Based on our experience with one loop diagrams we now propose an RG
action for the system of the form: 
\begin{widetext}
\begin{equation}
S\left(h\right)=-\int d^{d}\mathbf{x}dt\left[A\left(\frac{\partial h}{\partial t}\right)^{2}+B\left(\nabla h\right)^{2}+C\frac{\partial h}{\partial t}\left(\nabla h\right)^{2}+D\left(\frac{\partial h}{\partial t}\right)^{3}+E\left(\nabla h\right)^{4}+F\left(\nabla h\right)^{2}\left(\frac{\partial h}{\partial t}\right)^{2}+G\left(\frac{\partial h}{\partial t}\right)^{4}\right]\label{eq:RG_action}
\end{equation}
\end{widetext}

Here $A,B,C,D,E,F,G$ are coupling constants. All these terms will
be generated under perturbative RG and are the most RG relevant ones.
However in view of Ref. \citep{Goldstein_2024} we now mention the
limited reliability of obtaining specific values of the constants
$A,B,C,D,E,F,G$ (say at a fixed point) due to various forms of the
cutoff (which is so far unknown \citep{Goldstein_2024}). As such,
we do not do full RG analysis of this action as we will not reliably
obtain the values of the coupling coefficients $A,B,C,D,E,F,G$ and
leave it for future work.

\section{Conclusion}\label{sec:Conclusion}

In this work we started with the MSR based quantum action for the
KPZ equation and integrated out the Lagrange multiplier introduced
in the MSR procedure exactly to obtain a classical action e.g. an
action purely for the height field in Eq. (\ref{eq:Action_height})
which is the main result of the work. We did gaussian and one loop
analysis on the action in Eq. (\ref{eq:Action_height}). The gaussian
action led us to conclude that surfaces form smoothly for $d>2$.
The one loop analysis also predicts that $d>1$ is sufficient for
smooth surface growth. We had used RG regularization to get rid of
non-derivative (non-analytic) interactions. This is somewhat similar
to Hertz-Millis theory \citep{Continentino_2017,Hertz_1976,Millis_1993}
however can be regularized. However there are also some other concerns
with the RG approach in this situation \citep{Goldstein_2024} so
we did not pursue this further, due to some recent results \citet{Goldstein_2024}.
We did propose a general RG like action for KPZ (see Eq. (\ref{eq:RG_action})
where additional relevant terms will be generated perturbatively.
In the future it would be of interest to study the action in Eqs.
(\ref{eq:Action_height}) and (\ref{eq:RG_action}) much more thoroughly
in order to understand surface smoothness and roughness \citep{Kardar_1986,Kamenev_2011}
as well as to study possible applications to catalysis on surfaces
\citep{Cherendoff_2003}.

\textbf{Acknowledgements:} We would like to thank Davide Venturelli
for useful discussions.

\appendix

\part*{Supplementary online information}

\section*{KPZ with multiplicative noise}\label{sec:KPZ-with-multiplicative}

\subsection{Main idea}\label{subsec:Main-idea}

Here we extend the KPZ equations by claiming that due to geometric
considerations the noise $\xi$ should couple to the gradient the
same way as the main molecule deposition rate $\Gamma$ does. As such
we obtain an equation with multiplicatively coupled noise:
\begin{align}
\frac{\partial h}{\partial t} & =\left(\Gamma+\xi\right)\left(\sqrt{1+\left(\nabla h\right)^{2}}\right)-D\nabla^{2}h\nonumber \\
 & \cong\left(\Gamma+\xi\right)\left(1+\frac{1}{2}\left(\nabla h\right)^{2}\right)-D\nabla^{2}h\label{eq:KPZ_new}
\end{align}
We perform a MSR analysis \citep{Kamenev_2011} on this stochastic
differential equation and use the prescription in \citep{Kamenev_2011}
for the multiplicative noise in the system. We note that this approximation
means that most of the noise comes because of the deposition term
while the usual approximation means that most of the noise comes because
of the diffusion term. A more general approach (that incorporates
both limits) is given below.

\subsection{ Equation of motion for two sources of noise}\label{subsec:Computation-of-the}

We know that the equation of motion can be written as 
\begin{equation}
\frac{\partial h}{\partial t}\cong\left(\Gamma+\xi_{1}\right)\left(1+\frac{1}{2}\left(\nabla h\right)^{2}\right)-D\nabla^{2}h+\xi_{2}\label{eq:Heigh_general}
\end{equation}
Where we have that $\xi_{1}\left(\mathbf{x},t\right)$ is a Gaussian
noise function with action:
\begin{equation}
\int\mathcal{D}\xi_{1}\exp\left(-\int d^{d}\mathbf{x}dt\xi_{1}^{2}/2\Delta_{1}\right)\label{eq:Action-1}
\end{equation}
which corresponds to molecule deposition noise. While $\xi_{2}\left(\mathbf{x},t\right)$
is Gaussian white noise with action: 
\begin{equation}
\int\mathcal{D}\xi_{2}\exp\left(-\int d^{d}\mathbf{x}dt\xi_{2}^{2}/2\Delta_{2}\right)\label{eq:Gaussian_noise}
\end{equation}
Where \citep{Kamenev_2011}:
\begin{equation}
\frac{1}{\Delta_{2}}=\frac{\left(k_{B}T\right)^{2}}{2D}\label{eq:Fluctuation_dissipation_relation}
\end{equation}
which corresponds to molecule diffusion noise.

\subsection{ Action for KPZ equation with multiplicative noise}\label{sec:Main-idea}
\begin{widetext}
Now we write 
\begin{equation}
P\left(h\right)=\int\mathcal{D}\xi_{1}\mathcal{D}\xi_{2}\exp\left(-\int d^{d}\mathbf{x}dt\left[\xi_{1}^{2}/2\Delta_{1}+\xi_{2}^{2}/2\Delta_{2}\right]\right)\delta\left(\frac{\partial h}{\partial t}-\left(\Gamma+\xi_{1}\right)\left(1+\frac{1}{2}\left(\nabla h\right)^{2}\right)+D\nabla^{2}h-\xi_{2}\right)\label{eq:Two_noise_terms}
\end{equation}
We now study the stochastic differential equation in Eq. (\ref{eq:Two_noise_terms}).
We now perform MSR analysis \citep{Kamenev_2011} on this stochastic
differential equation in Eq. (\ref{eq:KPZ_new}) and we do not introduce
Faddeev-Popov ghosts \citep{Peskin_1995} for the functional determinant
$\det\left[1+\frac{1}{2}\left(\nabla h\right)^{2}\right]$ as it has
been absorbed into regularization \citep{Kamenev_2011}and change
variables $h\rightarrow h-\Gamma t$: 
\begin{align}
\mathcal{Z} & =\int\mathcal{D}h\mathcal{D}\xi_{1}\mathcal{D}\xi_{2}\exp\left(-\int d^{d}\mathbf{x}dt\left[\xi_{1}^{2}/2\Delta_{1}+\xi_{2}^{2}/2\Delta_{2}\right]\right)\delta\left(\frac{\partial h}{\partial t}-\left(\Gamma+\xi_{1}\right)\left(1+\frac{1}{2}\left(\nabla h\right)^{2}\right)+D\nabla^{2}h-\xi_{2}\right)\nonumber \\
 & =\int\mathcal{D}\xi_{1}\mathcal{D}\xi_{2}\mathcal{D}h\mathcal{D}p\exp\left(-\int d^{d}\mathbf{x}dt\left[\left[\xi_{1}^{2}/2\Delta_{1}+\xi_{2}^{2}/2\Delta_{2}\right]+p\left(\frac{\partial h}{\partial t}-\left(\Gamma+\xi_{1}\right)\left(1+\frac{1}{2}\left(\nabla h\right)^{2}\right)+D\nabla^{2}h-\xi_{2}\right)\right]\right)\nonumber \\
 & =\int\mathcal{D}\xi\mathcal{D}h\mathcal{D}p\exp\left(-\int d^{d}\mathbf{x}dt\left[\frac{1}{2\Delta_{1}}\left(\xi_{1}-\Delta_{1}p\right)^{2}+\frac{1}{2\Delta_{2}}\left(\xi_{2}-\Delta_{2}p\right)^{2}+\frac{1}{2}\xi_{1}p\left(\nabla h\right)^{2}+\right.\right.\nonumber \\
 & \qquad\qquad\qquad\left.\left.+p\left(\frac{\partial h}{\partial t}-\frac{1}{2}\Gamma\left(\nabla h\right)^{2}+D\nabla^{2}h-\left(\Delta_{1}+\Delta_{2}\right)p\right)\right]\right)\label{eq:Path_Integral_manipulations_I}
\end{align}
Whereby:
\begin{align}
\mathcal{Z} & =\int\mathcal{D}\xi\mathcal{D}h\mathcal{D}p\exp\left(-\int d^{d}\mathbf{x}dt\left[\frac{1}{2\Delta_{1}}\left(\xi_{1}-\Delta_{1}p\right)^{2}+\frac{1}{2\Delta_{2}}\left(\xi_{2}-\Delta_{2}p\right)^{2}+\frac{1}{2}\left(\xi_{1}-\Delta_{1}p\right)p\left(\nabla h\right)^{2}+\frac{1}{2}\Delta_{1}p^{2}\left(\nabla h\right)^{2}\right.\right.\nonumber \\
 & \qquad\qquad\qquad\left.\left.+p\left(\frac{\partial h}{\partial t}-\frac{1}{2}\Gamma\left(\nabla h\right)^{2}+D\nabla^{2}h-\left(\Delta_{1}+\Delta_{2}\right)p\right)\right]\right)\nonumber \\
 & =\int\mathcal{D}\Xi_{1}\mathcal{D}\Xi_{2}\mathcal{D}h\mathcal{D}p\exp\left(-\int d^{d}\mathbf{x}dt\left[\frac{1}{2\Delta_{1}}\Xi_{1}^{2}+\frac{1}{2\Delta_{1}}\Xi_{2}^{2}+\frac{1}{2}\Xi_{1}p\left(\nabla h\right)^{2}+\frac{1}{2}\Delta_{1}p^{2}\left(\nabla h\right)^{2}+\right.\right.\nonumber \\
 & \qquad\qquad\qquad\left.\left.+p\left(\frac{\partial h}{\partial t}-\frac{1}{2}\Gamma\left(\nabla h\right)^{2}+D\nabla^{2}h-\left(\Delta_{1}+\Delta_{2}\right)p\right)\right]\right)\nonumber \\
 & =\int\mathcal{D}h\mathcal{D}p\exp\left(-\int d^{d}\mathbf{x}dt\left[\frac{1}{4\Delta_{1}}p^{2}\left(\nabla h\right)^{4}+\frac{1}{2}\Delta_{1}p^{2}\left(\nabla h\right)^{2}+p\left(\frac{\partial h}{\partial t}-\frac{1}{2}\Gamma\left(\nabla h\right)^{2}+D\nabla^{2}h-\left(\Delta_{1}+\Delta_{2}\right)p\right)\right]\right)\label{eq:Path_Integral_Manipulations_II}
\end{align}
Where we have performed a change of variables $\mathcal{D}\xi_{1,2}\mathcal{D}h\mathcal{D}p\rightarrow\mathcal{D}\Xi_{1,2}\mathcal{D}h\mathcal{D}p$
with $\Xi_{1,2}=\xi_{1,2}-\Delta_{1,2}p$. We note that the determinant
of the Jacobian of the transformation being one as the Jacobian is
upper triangular and integrated out $\Xi$ to tree level (which is
exact as there is only one connected Feynman diagram involved (then
one can use exponentiation of disconnected diagrams)) to obtain the
action: 
\begin{equation}
S\left(h,p\right)=-\int d^{d}\mathbf{x}dt\left[\Delta_{1}p^{2}\left(\nabla h\right)^{4}+\frac{1}{2}\Delta_{1}p^{2}\left(\nabla h\right)^{2}+p\left(\frac{\partial h}{\partial t}-\frac{1}{2}\Gamma\left(\nabla h\right)^{2}+D\nabla^{2}h-\left(\Delta_{1}+\Delta_{2}\right)p\right)\right]\label{eq:MSR_action}
\end{equation}
Note that we used the same prescription as in Ref \citep{Kamenev_2011}
for multiplicative noise.
\end{widetext}

\end{document}